\documentclass[aps,prl,twocolumn,showpacs]{revtex4-1}
\usepackage{amsmath}
\usepackage[dvips,xdvi]{graphicx}
\usepackage{amssymb}
\usepackage{epsfig}

\begin{document}
\title{Lattice-induced rapid formation of spin singlets in spin-1 spinor condensates}
\author{L. Zhao}
\author{T. Tang}
\author{Z. Chen}
\author{Y. Liu}
\email{Electronic address: yingmei.liu@okstate.edu}
\affiliation{Department of Physics, Oklahoma State University,
Stillwater, Oklahoma 74078, USA}
\date{\today}
\begin{abstract}
We experimentally demonstrate that combining a cubic optical lattice with a spinor Bose-Einstein condensate substantially
relaxes three strict constraints and brings spin singlets of ultracold spin-1 atoms into experimentally accessible
regions. About 80 percent of atoms in the lattice-confined spin-1 spinor condensate are found to form spin singlets,
immediately after the atoms cross first-order superfluid to Mott-insulator phase transitions in a microwave dressing
field. A phenomenological model is also introduced to well describe our observations without adjustable parameters.

\end{abstract}


\maketitle

Many-body spin singlet states, in which multiple spin components of zero total spin are naturally entangled, have been
widely suggested as ideal candidates in investigating quantum metrology and quantum
memories~\cite{Ashhab2,Huang,Toth,Mueller,Mitchell2014,Ciccarello,Urizar-Lanz,Superlattice,Ho,Gerbier2016,Gerbier2013,Law,Eckert,Toth2}.
Advantages of spin singlets in the quantum information research include long lifetimes and enhanced tolerance to
environmental noises~\cite{Toth,Superlattice}. These advantages may become more pronounced if the singlets consist of
ultracold spin-1 particles~\cite{Huang}. A spin singlet is the ground state of many types of spinor gases, however, its
experimental realizations have proven to be very challenging mainly due to its
fragilities~\cite{Mueller,Ho,Gerbier2016,Gerbier2013,Hirano,Superlattice}. Allowed parameter ranges for spin singlets of
spin-1 atoms are strictly limited to the vicinity of zero quadratic Zeeman energy $q$ and zero magnetization $m$, and the
ranges drastically shrink when the atom number increases~\cite{Mueller,Law,Ho,Gerbier2016}. Another constraint is the
formation of spin singlets requires atoms remaining adiabatic for a long time duration~\cite{Gerbier2016, freespace}. In
this Letter, we experimentally demonstrate that combining a spinor Bose-Einstein condensate (BEC) with cubic optical
lattices significantly relaxes these strict constraints and enables creating spin singlets of spin-1 atoms rapidly. Our
observations confirm that spin singlets are brought into experimentally accessible regions by two key lattice-modified
parameters, which are the lattice-enhanced interatomic interactions and substantially reduced atom number in individual
lattice sites. Lattice-confined spinor BECs present degeneracies in spin and spatial domains, which provide perfect
platforms to simulate quantum mesoscopic systems and study rich physics of fragmentation~\cite{Ashhab2, Mueller}.

Different methods have been proposed for detecting spin singlets. The first approach is to measure the population of each
spin component, as atoms in a spin singlet should be evenly distributed into all spin states~\cite
{demler2003,Javanainen}. The second method is to verify a spin singlet is invariant after its spin is rotated by a
resonant Rf-pulse~\cite{Toth,Urizar-Lanz,Mueller,Javanainen,Zhou2003}. Another signature of a spin singlet is its high
level of spin squeezing shown in quantum non-demolition measurements~\cite{Toth,Eckert,Toth2,Mitchell2014}. A spin singlet
can also be identified by its high-order correlation functions, e.g., its zero spin nematicity detected by light
scattering measurements~\cite{Mueller,MuellerLightScattering}. Other detectable parameters of a spin singlet include large
population fluctuations in each of its spin components, and its excitation spectra mapped by Bragg
scattering~\cite{Ho,demler2003}. In this paper, we apply the first two methods to demonstrate that about 80$\%$ of spin-1
atoms in a lattice-confined spinor BEC can form spin singlets, immediately after the atoms cross first-order superfluid
(SF) to Mott-insulator (MI) phase transitions in a microwave dressing field. A phenomenological model is also developed to
explain our observations without adjustable parameters.
\begin{figure*}[t]
\includegraphics[width=176mm]{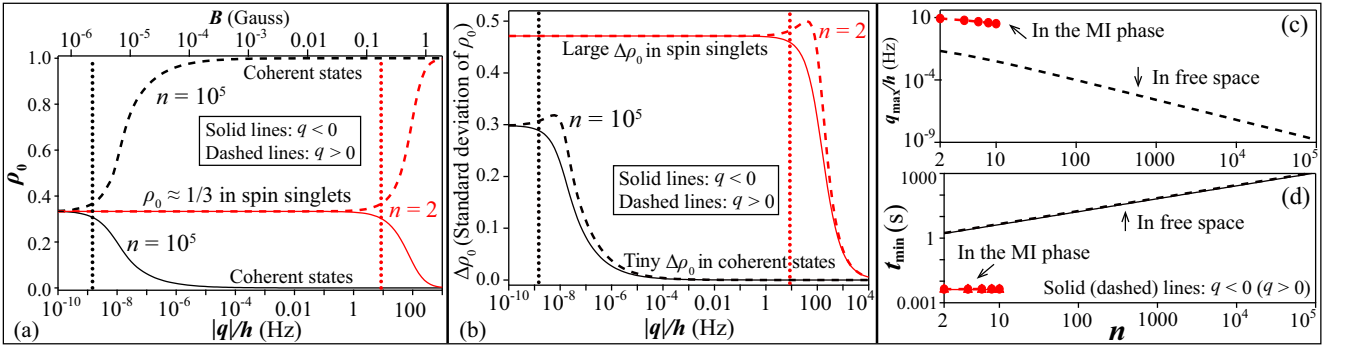}
\caption{(a) and (b): vertical black (red) dotted lines mark $q_{\rm max}$, the maximum allowed $q$ for spin singlets, in
$F$=1 sodium spinor BECs of $n=10^5$ atoms in free space (in the $n$=2 Mott lobe at $u_L=26E_R$). All panels are derived
from MFT at zero $m$ with solid (dashed) lines representing the $q<0$ ($q>0$) region, and black (red) lines representing
spinor gases in free space (spinor Mott insulators)~\cite{freespace}. (a) Predicted $\rho_0$ versus $|q|$ at $n=2$ (red)
and $10^5$ (black). The top horizontal axis lists the corresponding $B$ when $q>0$. (b) Predicted $\Delta\rho_0$ versus
$|q|$ at $n=2$ (red) and $10^5$ (black). (c) Predicted $q_{\rm max}$ versus $n$. (d) The minimum time $t_{\rm min}$ versus
$n$ for generating singlets of sodium atoms via an adiabatic sweep at its corresponding $\pm q_{\rm max}$.}
\label{figure1}
\end{figure*}

We start each experimental cycle with an antiferromagnetic $F$=1 spinor BEC of $n=1.2\times 10^5$ sodium atoms and zero
$m$ in its free-space ground state, i.e., a longitudinal polar (LP) state in the $q>0$ region or a transverse polar (TP)
state when $q<0$~\cite{JiangLattice,StamperKurnRMP,JiangGS,Zhao2dLattice,Ho1998Spinor}. The atoms are then loaded into
cubic lattices and enter into the MI phase with the peak occupation number per lattice site being five, $n_{\rm peak}=5$.
We express the Hamiltonian of the spinor Mott insulators by ignoring the hopping energy in the site-independent
Bose-Hubbard model as~\cite{JiangLattice}:\vspace{-1pt}
\begin{align}
\hat H = \dfrac{U_{0}}{2} (\hat n^{2}-\hat n) -\mu \hat n + \dfrac{U_{2}}{2} ({\hat{\textbf{\emph{S}}}}^2 -2\hat n)+ q
(\hat n_{1}+ \hat n_{-1})~.  \label{Eq:Ham}
\end{align}\vspace{-1pt}
Here $U_0$ ($U_2$) is the spin-independent (spin-dependent) interaction, $\mu$ is the chemical potential,
$\hat{\textbf{\emph{S}}}$ is the spin operator, and $\hat n=\sum_{m_F} \hat n_{m_F}$ is the number operator of all
hyperfine $m_F$ states. We obtain the ground states of spinor Mott insulators by diagonalizing Eq.~\eqref{Eq:Ham} at a
given $n$. For example, the ground states are spin singlets at zero $q$ in the even Mott lobes.

Sufficiently deep cubic lattices localize atoms and lower $n$ by five orders of magnitude in a typical BEC system.
Figure~\ref{figure1} illustrates how this enormous reduction in $n$ together with the lattice-enhanced interatomic
interactions can make spin singlets realizable in experimentally accessible regions. Figure~\ref{figure1} is derived from
the mean-field theory (MFT) and based on two notable signatures of a spin singlet, i.e., each of its $m_F$ states has an
identical fractional population $\rho_{m_F}$ and a big $\Delta\rho_{m_F}$ (the standard deviation of
$\rho_{m_F}$)~\cite{demler2003,Javanainen,Ho}. For example, spin singlets of $F$=1 atoms should have
$\rho_0\approx\rho_{\pm1}\approx1/3$ and $\Delta\rho_0=2\Delta\rho_{\pm1}>0.29$. In sharp contrast, $\rho_0=0$ and
$\rho_{\pm1}=0.5$ ($\rho_0=1$ and $\rho_{\pm1}=0$) with negligible $\Delta\rho_{m_F}$ are found in coherent TP (LP) states
when $q<0$ ($q>0$)~\cite{JiangGS}. The allowed $q$ range for spin singlets is $0\leq |q|\leq q_{\rm max}$, which is
determined by considering $\Delta\rho_{m_F}\gg 0$ and $\rho_0=(1+0.1)/3$ at $q=q_{\rm max}$ (that corresponds to
$\rho_0\simeq(1-0.1)/3$ at $q=-q_{\rm max}$) in MFT~\cite{qcriteria}. An expansion of ten orders of magnitude in $q_{\rm
max}$ is marked by vertical dotted lines in Figs.~\ref{figure1}(a) and \ref{figure1}(b), i.e., from a narrow region of
$|q|/h<2\times10^{-9}$~Hz in a free-space spinor BEC of $10^5$ atoms to a much broader range of $|q|/h<9$~Hz in $n$=2
spinor Mott insulators. Here $h$ is the Planck constant. This drastic raise in $q_{\rm max}$ as $n$ decreases is also
shown in Fig.~\ref{figure1}(c) for a wide range of achievable $n$. In addition, the lattice-induced big reduction in $n$
can relax the magnetization constraint on creating spin singlets by five orders of magnitude, because $|m|\lesssim0.15/n$
is required for singlets at zero $q$~\cite{mcriteria}. Figure~\ref{figure1}(d) indicates another big improvement made by
cubic lattices: $t_{\rm min}$ can be dramatically decreased by three orders of magnitude after a free-space spinor BEC
enters the MI phase~\cite{freespace}. Here $t_{\rm min}$ is the minimum time for generating singlets via adiabatically
sweeping one parameter, such as $q$ and the lattice depth $u_L$. Spin singlets of $F$=1 atoms can thus be created in
realistic experimental setups, e.g., in the spinor Mott insulators of $|m|\leq 0.05$ as confirmed by our experimental data
in Figs.~\ref{figure3} and \ref{figure4}.

In each experimental cycle, we prepare a LP or TP state at $q/h=40$~Hz by pumping all atoms in the undesired $m_F$ states
of a $F$=1 spinor BEC to the $F$=2 state with resonant microwave pulses, and blasting away these $F$=2 atoms via a
resonant laser pulse. We then quench $q$ to a proper value in microwave dressing fields~\cite{ZhaoUwave}, and load atoms
into a cubic lattice constructed by three standing waves along orthogonal directions. The lattice spacing is $532\,$nm,
while lattice beams are originated from a single-mode laser at $1064\,$nm and frequency-shifted by $20\,$MHz with respect
to each other. We use Kapitza-Dirac diffraction patterns to calibrate $u_L$. Each data point in this paper is collected
after atoms being abruptly released from a lattice at a fixed $u_L$ and expanding ballistically within a given time of
flight $t_{\rm TOF}$. The standard Stern-Gerlach absorption imaging is a good method to measure $\rho_{m_F}$ of spinor
gases in the SF phase. Stern-Gerlach separations become indiscernible, when atoms completely lose phase coherence in the
MI phase and the signal-to-noise ratio diminishes in TOF images. To measure $\rho_0$ in spinor Mott insulators, we develop
a two-step microwave imaging method as follows: 1) count the $m_{F}$=0 atoms with the first imaging pulse preceded by
transferring all atoms in the $|F=1,m_F=0 \rangle$ state to the $F$=2 state; 2) count all remaining atoms that are in the
$ m_{F}=\pm 1$ states with the second imaging pulse. We compare these two imaging methods using a free-space spinor BEC,
and find they give similar $\rho_0$ with a negligible difference (unless specified, all quoted uncertainties are 2
standard errors).
\begin{figure}[tb]
\includegraphics[width=85mm]{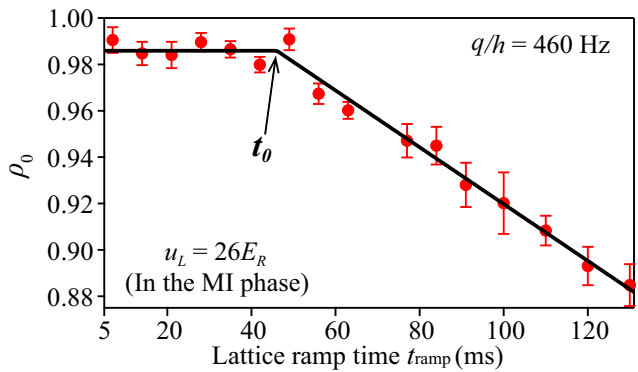}
\caption{Measured $\rho_0$ versus $t_{\rm ramp}$ after an initial LP spinor BEC enters the MI phase in a high field. Black
lines are two linear fits. We estimate $t_0$, the ideal $t_{\rm ramp}$, from the intersection point of these two lines
(see text).} \label{figure2}
\end{figure}

To ensure atoms adiabatically enter the MI phase, a cubic lattice is linearly ramped up within time $t_{\rm ramp}$ to
$u_L=26E_R$. Here $E_R$ is the recoil energy~\cite{Zhao2dLattice}. We carefully select $t_{\rm ramp}$ based on three
criteria. First, $t_{\rm ramp}$ should be long enough to satisfy $d u_L/dt\ll 32\pi E_R^2/h$, the interband adiabaticity
requirement~\cite{Ketterle2006}. Second, $t_{\rm ramp}$ should be larger than the MFT predicted $t_{\rm min}$, as
explained in Fig.~\ref{figure1}(d). These two criteria set $t_{\rm ramp}>5$~ms for our system. On the other hand, $t_{\rm
ramp}$ should be sufficiently short, with $t_{\rm ramp}\leq t_0$ to ensure lattice-induced heating is negligible and atom
losses are not greater than $10\%$. Figure~\ref{figure2} explains how we determine $t_{0}$ from the observed relationship
between $t_{\rm ramp}$ and $\rho_0$ in spinor Mott insulators at $u_L=26 E_R$ and $q/h=460$~Hz. In such a high field,
SF-MI phase transitions are second order because $U_{2}=0.04U_{0}>0$ and $q\gg U_{2}$ at this $u_L$ for the sodium
atoms~\cite{JiangLattice}. Atoms initially in a LP state should thus stay in the LP state with $\rho_0\simeq1$, as they
adiabatically cross the phase transitions and enter into the MI phase~\cite{JiangLattice}. The value of $\rho_0$ quickly
drops when inevitable heating is induced by lattices in a non-adiabatic lattice ramp sequence. We extract $t_{0}$ from the
intersection point of two linear fits to the data in Fig.~\ref{figure2}, which yields $t_{\rm ramp}\leq t_0\approx 45$~ms.
Within this acceptable $t_{\rm ramp}$ range, a slower lattice ramp is preferred because it could more easily keep the
system adiabatic and provide sufficient time for the atom redistribution processes~\cite{redistribution}. The ideal
lattice ramp speed is therefore set at $d u_L/dt= 26E_R/t_0$ for our system.

The opposite limit is $|q|\ll U_2$ near zero $q$, where SF-MI phase transitions are first order and spin singlets are the
ground state of $F$=1 spinor gases in the even Mott lobes~\cite{JiangLattice}. We may thus identify the formation of spin
singlets from evolutions of $\rho_0$ and $\Delta\rho_0$ during a first-order SF-MI transition. Figure~\ref{figure3} shows
two such evolutions when atoms initially in the TP state are adiabatically loaded into the cubic lattice at the ideal
lattice ramp speed to various final $u_L$ in $q/h = -4~\rm{Hz}$. These evolutions have three distinct regions. In the SF
phase where $0\leq u_L\leq 15E_R$, atoms remain in the TP state with $\rho_0=0$ and negligible $\Delta\rho_0$. As atoms
cross first-order SF-MI transitions in $15E_R\leq u_L\leq18E_R$, $\rho_0$ and $\Delta\rho_0$ sigmoidally increase with
$u_L$. When all atoms enter into the MI phase at $u_L\geq 21E_R$, both $\rho_0$ and $\Delta\rho_0$ reach their equilibrium
values of $\rho_0\approx0.3$ and $\Delta\rho_0\gg 0$. These observations qualitatively agree with the characteristics of
spin singlets. Despite that other factors can also increase $\Delta\rho_0$ in the MI phase, the measured $\Delta\rho_0$ is
much smaller than the MFT prediction shown in Fig.~\ref{figure1}(b). This may be due to the fact that the observed
$\Delta\rho_0$ is an average over all $5\times10^4$ lattice sites in our system. Unless one can detect single lattice site
precisely, the value of $\Delta\rho_0$ may not be used to verify spin singlets in lattice-confined spinor gases. We also
monitor the time evolution of atoms at fixed $u_L$ and $q$ after the ideal lattice ramp sequence. No spin oscillations are
found at each $q$ studied in this paper, which confirms atoms always stay at their ground states in these ideal lattice
sequences.
\begin{figure}[tb]
\includegraphics[width=85mm]{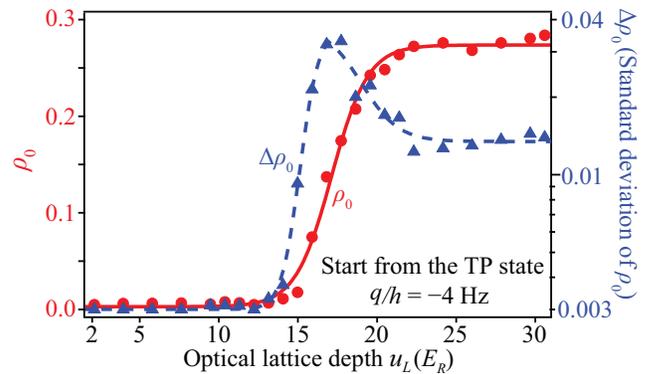}
\caption{Measured $\rho_0$ (red circles) and $\Delta\rho_0$ (blue triangles) versus $u_L$ after an initial TP spinor BEC
undergoes the ideal lattice sequence to various final $u_L$ in a weak field near zero $q$. The solid line is a sigmoidal
fit, and the dashed line is to guide the eye.} \label{figure3}
\end{figure}

We observe similar $\rho_0$ and $\Delta\rho_0$ evolutions within a wide range of $q$ near zero field. The measured
$\rho_0$ versus $q$ in spinor Mott insulators at $u_L=26E_R$ is shown in Fig.~\ref{figure4}(a). These Mott insulators of
$n_{\rm peak}=5$ are inhomogeneous systems, in which $\rho_0$ at a fixed $q$ may be given by the weighted average over all
Mott lobes:\vspace{-1pt}
\begin{align}
\rho_0 =&\sum_{j=1}^5\rho_{0_j}\chi_j~.\label{eq:rhoMF}
\end{align}\vspace{-1pt}
Here $\rho_{0_j}$ is the MFT predicted $\rho_0$ in the ground state $\psi_j$ of the $n$=$j$ Mott lobe, and $\chi_{j}$
represents mean-field atom density distributions in a harmonic trap~\cite{JiangLattice}. The prediction of
Eq.~\eqref{eq:rhoMF} shown by red dashed lines in Fig.~\ref{figure4}(a), however, appears to largely disagree with our
data. To understand this big discrepancy, we have tried several models and found only one phenomenological model can
surprisingly describe our data without adjustable parameters (see black solid lines in Fig.~\ref{figure4}(a)). This
phenomenological model is based on one major difference between spinor and scalar Mott insulators predicted by the
Bose-Hubbard model: i.e., the formation of spin singlets enlarges even Mott lobes in antiferromagnetic spinor
gases~\cite{JiangLattice}. For example, the $n$=2 even Mott lobe emerges at $u_L\approx16.5E_R$, while the $n$=3 odd Mott
lobe only exists in a much deeper lattice of $u_L\geq19.5E_R$ for $F$=1 sodium spinor gases near zero
field~\cite{JiangLattice}. In the intermediate lattice depth of $16.5E_R<u_L<19.5E_R$ near zero $q$, atoms in the $n$=3
lattice sites can freely tunnel among adjacent lattice sites, while particles in an $n$=2 lattice site already enter into
the MI phase and are localized in this site. At a proper $u_L$ near zero $q$, atoms may thus be able to redistribute among
lattice sites with a given odd $n$ in the lattice-confined spinor gases. For example, at $u_L=19E_R>16.5E_R$, the
tunneling of one atom converts two adjacent $n$=3 lattice sites to one $n$=2 and one $n$=4 sites. This $u_L$ is then deep
enough to localize the six atoms by forming a two-body spin singlet in one site and a 4-body spin singlet in the other
site~\cite{redistribution}. As a result of similar redistribution processes, atoms initially in lattice sites with $n=5$
may form 4-body and 6-body spin singlets in the ideal lattice ramp sequences. In contrast, redistribution processes may
not occur among the $n$=1 lattice sites, because the $n$=1 and $n$=2 Mott lobes emerge at similar $u_L$ for the sodium
atoms. Our phenomenological model takes these atom redistribution processes into account, and expresses $\rho_0$ in the
spinor Mott insulators created by the ideal lattice ramp sequence as\vspace{-1pt}
\begin{align}
\rho_0 =&\sum\limits_{j=3,5}\chi_j\frac{(j+1)\rho_{0_{j+1}}+(j-1)\rho_{0_{j-1}}}{2j}\nonumber\\
&+\sum\limits_{j=1,2,4}\rho_{0_j}\chi_j~.\label{eq:rho}
\end{align}
Figure~\ref{figure4}(a) shows that the prediction of Eq.~\eqref{eq:rho} agrees with our experimental data. The validity of
this phenomenological model is also verified by comparing its prediction with the observed $\rho_0$, after a resonant
Rf-pulse is applied to rotate the spin of atoms by 90 degrees. In this paper, the spin rotation operator $\hat R_x=\exp(-i
\frac{\pi}{2} \hat S_x)$ is along the $x$-axis, which is orthogonal to the quantization axis ($z$-axis). After $\pi/2$
spin rotations, $ \rho_{0_j} $ in Eq.~\eqref{eq:rho} changes to $\rho_{0_j}^r=\frac{\langle\psi_j|\hat R^\dagger_x\hat n_0
\hat R_x|\psi_j\rangle}{\langle\psi_j|\hat R^\dagger_x \hat n \hat R_x|\psi_j\rangle}$ in the $n$=$j$ Mott lobe. The
prediction of Eq.~\eqref{eq:rho} after these spin rotations is shown by the upper black solid line in
Fig.~\ref{figure4}(a), which well agrees with our data. The two data sets in Fig.~\ref{figure4}(a) respectively represent
projections of the atomic spin along two orthogonal axes. The observed good agreements between our phenomenological model
and these data sets, therefore, suggest this model may reveal mechanisms of the ideal lattice ramp sequence in
antiferromagnetic spinor gases.
\begin{figure}[t]
\includegraphics[width=85mm]{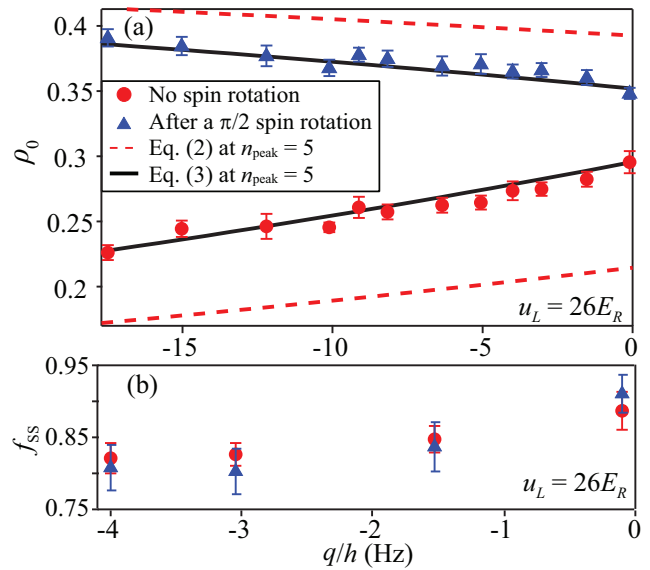}
\caption{(a) Red circles (blue triangles) are the measured $\rho_0$ in spinor Mott insulators without (with) atoms being
rotated by resonant $\pi/2$ pulses at various $q$. The black solid (red dashed) line is the prediction of Eq.~(3)
(Eq.~\eqref{eq:rhoMF}). (b) Spin singlet fraction $f_{\rm ss}$ extracted from Panel (a) versus $q$ (see text). The
insulators are created after an initial TP spinor BEC undergoes the ideal lattice ramp sequence.} \label{figure4}
\end{figure}

Our data taken with and without the $\pi/2$ spin rotations appear to converge to a value around $\rho_0\approx1/3$ as $q$
gets closer to zero in Fig.~\ref{figure4}(a). This indicates the spinor Mott insulators become more rotationally invariant
near zero field. As the spin rotational invariance is one unique signature of spin singlets, the reduced gap between the
two data sets in Fig.~\ref{figure4}(a) implies significant amounts of atoms may form spin singlets when $q$ approaches
zero. In our system, about $10\%$ of atoms stay in the $n$=1 Mott lobe where no spin singlet can be formed. This accounts
for the observed small gap between the two data sets near zero $q$ in Fig.~\ref{figure4}(a), and limits the maximum
$f_{\rm ss}$ realizable in our system to about $90\%$. Here $f_{\rm ss}$ represents the fraction of atoms forming spin
singlets in spinor gases. We extract $f_{\rm ss}$ from the measured $\rho_0 $ based on Ref.~\cite{fss}. The two data sets
in Fig.~\ref{figure4}(a) appear to yield similar $f_{\rm ss}$ at a fixed $q$ near zero field: i.e., $f_{\rm
ss}\approx80\%$ when $-4~{\rm{Hz}}\leq q/h\leq0~{\rm{Hz}}$ as shown in Fig.~\ref{figure4}(b). This indicates around $80\%$
of atoms form spin singlets in our system. Similar phenomena and slightly smaller $f_{\rm ss}$ are also observed in spinor
Mott insulators generated after atoms initially in the LP state cross first-order SF-MI transitions in the ideal lattice
ramp sequences when $q>0$.

In conclusion, our experimental data have confirmed that combining cubic lattices with spinor BECs makes spin singlets of
ultracold spin-1 atoms achievable in experimentally accessible regions. Via two independent detection methods, we have
demonstrated that about 80$\%$ of atoms in the lattice-confined $F$=1 spinor BEC form spin singlets, after the atoms cross
first-order SF-MI phase transitions near zero field. We have developed a phenomenological model that explains our
observations without adjustable parameters. Our recent work has also indicated that we may be able to identify another
signature of spin singlets, i.e., confirm their zero spin nematicity in light scattering measurements~\cite{ongoing}.
\begin{acknowledgments}
We thank the National Science Foundation and the Oklahoma Center for the Advancement of Science and Technology for
financial support.
\end{acknowledgments}


\begin{thebibliography}{100}
\bibitem{freespace} Our MFT calculations are based on Ref.~\cite{JiangGS} with $t_{\rm min}=h/\Delta E$. Here $\Delta E$ is the energy gap between the ground state and the first excited state.
\bibitem{mcriteria}
We derive this magnetization constraint from Eq.~3 of Ref.~\cite{Ho}.
\bibitem{redistribution} The required time for the atom redistribution process is proportional to $(6 n J)^{-1}$ in cubic lattices. This may set $t_{\rm ramp}\geq35$~ms for our system.
\bibitem{fss}
Our phenomenological model predicts that $\rho_0=f_{\rm ss}/3$ ($=f_{\rm ss}/3+(1-f_{\rm ss})/2$) in spinor Mott
insulators created from an initial TP spinor BEC in the ideal lattice sequence without (with) $\pi/2$ spin rotations. This
prediction only works near zero $q$ and slightly underestimates $f_{\rm ss}$.
\end{thebibliography}
\end{document}